\documentclass[
  journal=largetwo,
  manuscript=article-type,
  year=2020,
  volume=37,
]{cup-journal}

\usepackage{amsmath}
\usepackage[nopatch]{microtype}
\usepackage{booktabs}
\usepackage{subcaption}
\usepackage{graphicx} 

\title{Performance of the Segment Anything Model in Various RFI/Events Detection in Radio Astronomy}

\author{Yanbin Yang}
\affiliation{Shanghai Astronomical Observatory, Chinese Academy of Sciences, Shanghai 200030, China}
\alsoaffiliation{University of Chinese Academy of Sciences, Beijing 101408, China}
\alsoaffiliation{School of Physical Science and Technology, ShanghaiTech University, Shanghai 201210, China}

\author{Feiyu Zhao}
\affiliation{Shanghai Astronomical Observatory, Chinese Academy of Sciences, Shanghai 200030, China}
\alsoaffiliation{University of Chinese Academy of Sciences, Beijing 101408, China}
\author{Ruxi Liang}
\affiliation{Shanghai Astronomical Observatory, Chinese Academy of Sciences, Shanghai 200030, China}

\author{Quan Guo}
\affiliation{Shanghai Astronomical Observatory, Chinese Academy of Sciences, Shanghai 200030, China}
\alsoaffiliation{Key Laboratory of Radio Astronomy and Technology, Chinese Academy of Sciences, Beijing 100101, China}
\email[Quan Guo \& Junhua Gu]{guoquan@shao.ac.cn, jhgu@nao.cas.cn}

\author{Junhua Gu}
\affiliation{National Astronomical Observatories, Chinese Academy of Sciences, 20A Datun Road, Beijing 100101, China}
\author{Yan Huang}
\affiliation{National Astronomical Observatories, Chinese Academy of Sciences, 20A Datun Road, Beijing 100101, China}

\author{Yun Yu}
\affiliation{Shanghai Astronomical Observatory, Chinese Academy of Sciences, Shanghai 200030, China}

\keywords{methods: data analysis, methods: observational, techniques: image processing} %% First letter not capped

\begin{document}

\begin{abstract}
The emerging era of big data in radio astronomy demands more efficient and higher-quality processing of observational data. While deep learning methods have been applied to tasks such as automatic radio frequency interference (RFI) detection, these methods often face limitations, including dependence on training data and poor generalization, which are also common issues in other deep learning applications within astronomy. In this study, we investigate the use of the open-source image recognition and segmentation model, Segment Anything Model (SAM), and its optimized version, HQ-SAM, due to their impressive generalization capabilities. We evaluate these models across various tasks, including RFI detection and solar radio burst (SRB) identification. For RFI detection, HQ-SAM (SAM) shows performance that is comparable to or even superior to the SumThreshold method, especially with large-area broadband RFI data. In the search for SRBs, HQ-SAM demonstrates strong recognition abilities for Type II and Type III bursts. Overall, with its impressive generalization capability, SAM (HQ-SAM) can be a promising candidate for further optimization and application in RFI and event detection tasks in radio astronomy. 
\end{abstract}

\section{Introduction}

In recent years, significant advances have been made in radio astronomy. The Square Kilometre Array (SKA, \citealt{2009IEEEP..97.1482D}), the world's largest radio telescope currently under construction, is expected to deliver groundbreaking discoveries while addressing the challenge of handling the vast amounts of data it will generate. Meanwhile, many exciting scientific endeavors, such as understanding the cosmic dawn (CD) and the epoch of reionization (EoR) by observing the ultra-faint neutral hydrogen redshifted 21-cm line signal at low frequencies, will require exceptionally high-quality data, minimizing RFI (\citealt{rfimethod, Offringa_2010}) as much as possible. There is evidence suggesting that even fainter RFIs in Murchison Widefield Array (MWA, \citealt{MWA}) data are likely to contaminate the EoR power spectrum (\citealt{10.48550/arxiv.2310.03851}). The need for larger and higher-quality data sets drives the development of automatic RFI mitigation methods to reduce the manual burden and increase the efficiency and precision of the results.
 
Generally, radio frequency interference is different from celestial radio signals, which will contaminate the astronomical data and severely affect the observation results of radio telescopes (\citealt{rfimethod}). The sources of RFI are diverse, including radio broadcasts (\citealt{Huang_201621cma}), digital television (DTV, \citealt{Wilensky_2019}), satellites (\citealt{Sokoowski2015TheSO}), meteor trails (\citealt{Zhao_2023}), lightning (\citealt{Sokoowski2015TheSO}), aircraft communication (\citealt{2023transient}), and internal sources such as computers and screens (\citealt{Porko2011RadioFI}). Due to the continuous expansion of human activities and the rapid development of radio technology, radio telescopes are facing an increasing amount of RFI. Mitigating the RFI has become increasingly challenging.With the deluge of big data from new radio telescopes coming into operation and the need for high-quality data, the mitigation of RFI must be considered a pressing and crucial issue that must be addressed (\citealt{2023enh}).

The complex temporal and frequency structures of RFI make it challenging to model and obtain a universal method of RFI mitigation (\citealt{rfimethod}). In fact, many different methods have been proposed and adopted based on specific circumstances. In the proactive mitigation stage, there are approaches such as establishing a radio quiet zone (RQZ, \citealt{qitai}), setting up protected frequency bands (\citealt{Furlanetto:2006}), choosing locations with topographical features such as mountains (\citealt{mountain}), and using multi-layer electromagnetic shielding (\citealt{qitai, elec}). It is still necessary to adopt reactive mitigation methods in subsequent steps. \citet{Offringa_2010} proposed a combinatorial threshold algorithm: the SumThreshold method. AOFlagger, developed based on this threshold method, has been applied in Low Frequency Array (LOFAR) and MWA (\citealt{Offringa_2012, offringa2015}). \citet{Wilensky_2019} introduced a new method called Sky-Subtracted Incoherent Noise Spectra (SSINS), and its effectiveness was demonstrated by applying it to several kinds of RFI identified in data from MWA. For the 21 CentiMeter Array (21CMA), time-varying RFI is mitigated by handling weighted visibilities (\citealt{Huang_201621cma, antao}).

With rapid advancements in machine learning and deep learning, an increasing number of studies are applying machine learning to the field of astronomy in the face of vast observational data. In research on the mitigation of RFI, a large number of models based on deep learning have been used to detect RFI. \citet{cnn-sun} proposed a model using the convolutional neural network (CNN) to identify RFI. \citet{unet} successfully developed the U-Net in the field of RFI detection. To address the frequent errors encountered when CNN is used for the RFI flagging of FAST data, RFI-net was developed (\citealt{Yang_2020}). The mask region-based convolutional neural network (mask R-CNN) has been integrated with point-based rendering (PointRend) to identify RFI when finding HI galaxies (\citealt{liang2023detecting}).

Compared with traditional methods that require manual intervention to specify algorithm parameters, deep learning greatly improves the efficiency of data processing. However, it also has certain limitations. Model training, for instance, requires that a large amount of RFI data to be prepared as a training set, which is a burden. Using simulated RFI data to train models is a common choice. However, considering that real-world RFIs are often much more complex than mock data, models trained on simple simulated RFI may not perform optimally for complicated RFI recognition (\citealt{Yang_2020}). Additionally, there is a risk that the model may become dependent on training data or overfit, leading to missed detections of unknown RFI. Existing models are usually designed for specific tasks or telescopes and often lack sufficient generalization capability, making them cumbersome to use for other applications.The same issues arise in other fields where deep learning is applied, such as SRB detection and the search for pulsars or FRBs. These applications also suffer from problems like too few samples to build a training set and data imbalance (\citealt{SRB, pulsar_liu}). Thus, we wonder if it is possible to develop a method based on an open-source model with strong generalization capability that researchers from various fields can use directly with minimal fine-tuning or structural adjustment. This approach would not only reduce the burden of designing and training models from scratch but also minimize the drawbacks associated with model training, while hopefully retaining the ability to identify unknown events.

\citet{kirillov2023segment} proposed a model called the Segment Anything Model (SAM) for image recognition and segmentation. They utilized model-in-the-loop dataset annotation to construct the largest segmentation dataset to date, containing over 1 billion masks on 11 million images, for training. This approach gave SAM powerful generalization and zero-shot capability, which has been proven by its application in various fields (\citealt{samuse1, samuse2}). HQ-SAM builds on SAM by adding a High-Quality Output Token to improve the quality of the predicted masks while preserving SAM's generalization capability (\citealt{hqsam}). Inspired by the impressive generalization capability and ease of use of HQ-SAM (SAM), in this paper, we apply it to the field of radio astronomy. We explore its performance in RFI or events detection such as SRB, demonstrating the model's good generalization capability and recognition ability in this domain. By using the HQ-SAM (SAM) model, along with minor fine-tuning and modifications, we hope to find a better solution to the aforementioned challenges.

The structure of this paper is as follows. Section~\ref{sect:method} introduces SAM, HQ-SAM, and the SumThreshold method, which we use for comparison with the first two in detail. Section~\ref{sect: real RFI} shows the real RFI from the 21CMA and the detection results using the three methods. Section~\ref{sect:RFI simulation} demonstrates the recognition results of our mock RFI. Section~\ref{sect:SRB} applies HQ-SAM to the search for SRB. We discuss our findings in Section~\ref{disccussion} and provide our conclusions in Section~\ref{sect:conclusion}.

\section{Method}
\label{sect:method}

In this section, we will introduce several methods and techniques used in subsequent research, including SAM, HQ-SAM, and the SumThreshold method.

\subsection{Segment Anything Model (SAM) and HQ-SAM}

The Segment Anything Model is a groundbreaking image recognition and segmentation model capable of generating valid segmentation masks when provided with any form of segmentation prompt (points, box, mask, text, etc.). SAM consists of three components: an image encoder, a prompt encoder, and a lightweight mask decoder that combines the information from the first two to predict segmentation masks (\citealt{kirillov2023segment}). When using SAM, one simply needs to provide prompts indicating the content to be segmented in the image. SAM offers an automatic segmentation mode that evenly distributes points on the segmentation image to act as point prompts. The output will include multiple masks due to SAM's ambiguity-aware feature. Because the researchers built a massive segmentation dataset named SA-1B, containing over 1 billion masks and 11 million images to train the model, SAM has strong zero-shot capability (meaning the model can successfully identify or segment objects it has never specifically been trained on) and demonstrates outstanding performance across many tasks.

Due to the impressive zero-shot and generalization capability of SAM, it has been applied to a wide range of fields. \citet{MedSAM} proposed MedSAM by fine-tuning SAM with a medical image training dataset, demonstrating that this model can produce accurate segmentation in various medical image segmentation tasks. RSPrompter (\citealt{rsprompter}) is a method that creates appropriate prompts for SAM, enabling SAM to perform well in instance segmentation tasks for remote sensing images. \citet{cropsambased} utilized two pre-trained object detection models, You Only Look Once (YOLO)-v8 and DETR with Improved deNoising anchOr boxes (DINO), to first detect objects in the image and then used the detected bounding boxes as box prompts for SAM. This approach allowed SAM to achieve good scores in panoptic segmentation tasks for weeds and crops.

Although SAM has shown promising results in various fields, there are still some limitations in its application. SAM cannot automatically interpret images from different domains to generate appropriate prompts for itself or provide semantic categories for the predicted masks (\citealt{matcher}). As a result, many works design additional components to automatically generate suitable prompts when applying SAM, such as \citet{rsprompter} and \citet{cropsambased}, which were mentioned earlier. Additionally, SAM exhibits certain shortcomings when dealing with targets featuring complex background interference, ambiguous boundaries, or low image contrast (\citealt{medical}).

\citet{hqsam} identified two key issues with the segmentation results of SAM in some cases: coarse mask boundaries and incorrect predictions, which can significantly affect SAM's applicability and effectiveness. To address these problems, they proposed an upgraded model called HQ-SAM\footnote{https://github.com/SysCV/sam-hq}. HQ-SAM enhances the quality of predicted masks by incorporating a learnable High-Quality Output Token into SAM's mask decoder. This method allows HQ-SAM to retain the pre-trained model weights of SAM, thus preserving SAM's original zero-shot capability, while also enabling more precise segmentation across various tasks.

Overall, there is no significant difference between running HQ-SAM and SAM. SAM resizes input images of any size to 1024$\times$1024 pixels, meaning that smaller images will be enlarged more, which facilities better features extraction by the model. However, there is a potential risk of image quality loss with this enlargement, so it is important to ensure the clarity and detail of the input images. Considering the above points, users need to select the size best suited to their specific task. In automatic segmentation mode, SAM provides several adjustable parameters to optimize performance. For example, the parameter $points\_per\_side$ determines the number of sampling points along one side of the image, with the total number of points being $points\_per\_side^2$. Generally, increasing the number of sampling points improves recognition accuracy but also demands longer processing times, necessitating a trade-off. For more details on these parameters, please refer to SAM's documentation\footnote{https://github.com/facebookresearch/segment-anything}.

In this work, our primary focus is on applying SAM (HQ-SAM) to astronomical research areas including RFI and SRB, where deep learning techniques can offer significant benefits. We aim to assess whether we can achieve a broadly applicable model in astronomy with minimal cost (i.e., without extensive fine-tuning or structural modifications) while mitigating common issues associated with model training. To this end, we utilize the automatic segmentation mode provided by SAM, employing the same parameters for both SAM and HQ-SAM. Except for images of SRBs, which are sized at 1225$\times$645 pixels, all other images are standardized to 200$\times$200 pixels so that it will be enlarged by about 25 times, allowing for better recognition results. 

\subsection{The SumThreshold Method}

In the post-correlation stage (i.e., the stage following the correlation of signals) of RFI mitigation, thresholding is an effective method for removing strong RFI. The VarThreshold method, a combinatorial thresholding technique, iteratively combines samples and compares them with a strictly decreasing series of sample thresholds $\left \{ \chi _i \right \} ^N_{i=1}$, where  $N$ is the number of iteration. If the absolute values of all the samples exceed the threshold $\chi_{N}$, then these samples are flagged as RFI.

\citet{Offringa_2010} proposed the SumThreshold method, which optimizes the VarThreshold method. The SumThreshold method retains the content of the VarThreshold method regarding sample number selection and thresholds for each iteration. Unlike the VarThreshold method, the SumThreshold method calculates the sum of the statistics of $M$ samples and compares it with $M$ times the corresponding threshold $\chi_{N(M)}$. If it exceeds the threshold, all $M$ samples are flagged as contaminated. To prevent excessive false positives, the SumThreshold method also includes an additional condition: if a higher threshold has already flagged samples as RFI contaminated, the samples will be excluded from the summation in subsequent iterations, and their values will be replaced by the value of the current iteration's threshold. Compared to the VarThreshold method, the SumThreshold method allows the flagging of a sequence containing samples with values below the thresholds (\citealt{Offringa_2010}). Thus, it produces fewer false negatives, can flag weaker contaminations, and is more easily applied to various types of data. It also exhibits stronger robustness to abnormal values and noise in the samples. Refer to Appendix~\ref{sumthreshold appendix} for more details about the SumThreshold method.

The SumThreshold method has been widely applied in astronomical data processing, including the AOflagger pipeline for LOFAR \citep{offringa2010lofar} and an open-source Python package: Signal Extraction and Emission Kartographer (SEEK) \citep{seek}. However, the SumThreshold method also faces some challenges, such as the need for manual fine-tuning of thresholds or algorithm parameters in practical applications \citep{Yang_2020}, limitations in flagging RFI fainter than the single baseline thermal noise \citep{Wilensky_2019}, as well as being less effective in identifying broadband signals or extremely large amounts of RFI \citep{Yang_2020}. In our work, we use the SEEK\footnote{https://github.com/cosmo-ethz/seek} Python package to implement the SumThreshold method.

\section{RFI Detection of Real Data of the 21CMA}
\label{sect: real RFI}

In this section, we will demonstrate the results of applying the three methods, introduced in Section~\ref{sect:method} , on the real RFI of the 21CMA. 

\subsection{Observational Data of 21CMA}
 The 21CMA (\citealt{Zheng_201221cma}) is located at Ulastai, Xinjiang, China, which consists of 81 stations with a total of 10287 log-periodic antennas. It operates in the frequency range of 50-200 MHz, designed to detect the EoR. 

Our data is the same as in the work of \citet{gao}, which are the self-correlation spectra from two stations respectively: $E5$ (the 5th station in the east) and $E9$ (the 9th station in the east). The time resolution is about 1 ms, and the frequency range is 50-200 MHz. The observations were made on January 3, 4, and 5, 2021, with a total accumulated observation time of 42 hours and a total data volume of 4.6 TB \citep{gao}.

   \begin{figure}[hbt!]
   \centering
   \includegraphics[width=1\columnwidth, angle=0]{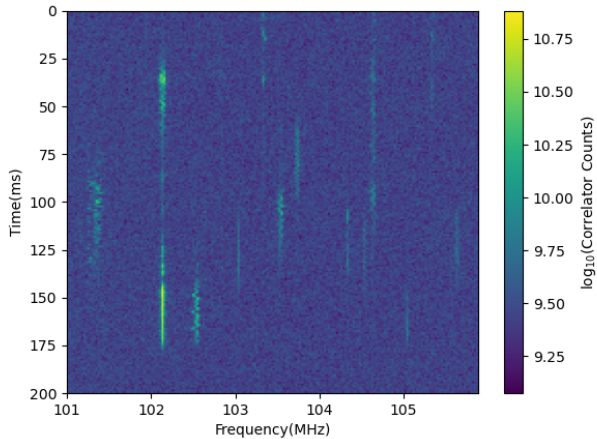}
   \caption{ An example of transient narrowband RFI (scattered FM radio signals) from the 21CMA is shown below. The horizontal axis represents frequency, while the vertical axis represents time. The image is 200$\times$200 in size.}
   \label{21CMARFI}
   \end{figure}

The raw spectrum data of the 21CMA is split into 8192 channels covering the frequency band 0~200 MHz. We cut the waterfall of the 21CMA data with 8192 pixels in frequency axis into stripes with 200 pixels to create images of size 200$\times$200. Fig.\ref{21CMARFI} presents an example of RFI from the 21CMA in the form of a waterfall plot. The horizontal axis represents the frequency, ranging from 101 MHz to approximately 106 MHz, while the vertical axis represents time, with a duration of 200 ms. The plot exhibits characteristics of intensity changes and fluctuations across multiple frequency channels over time, appearing more like compact polylines rather than straight lines. This pattern is a typical example of scattered frequency modulation (FM) radio signals. Within our observed frequency band, there are numerous remote FM broadcast signals, such as Urumqi Traffic Radio (97.4 MHz) and News Music Radio (99.0 MHz). These RFI may be attributed to the scattering of FM broadcasts by meteor trails or airplanes. In our real RFI detection, we also find long-duration continuous narrowband RFI corresponding to FM broadcasts. More details about the characteristics of RFI in the 21CMA will be discussed in Section~\ref{subsec:classification} .

\subsection{The Comparion of Detection Results by Different Methods}

Two waterfall plots are selected as representatives of continuous narrowband RFI and transient narrowband RFI, and the detection results of RFI are demonstrated. The horizontal axis of the waterfall plot represents frequency, spanning approximately 4.9 MHz of bandwidth, and the vertical axis represents time, with a duration of 200 ms (the subsequent waterfall plots follow the same specifications).

\begin{figure*}[hbt!]
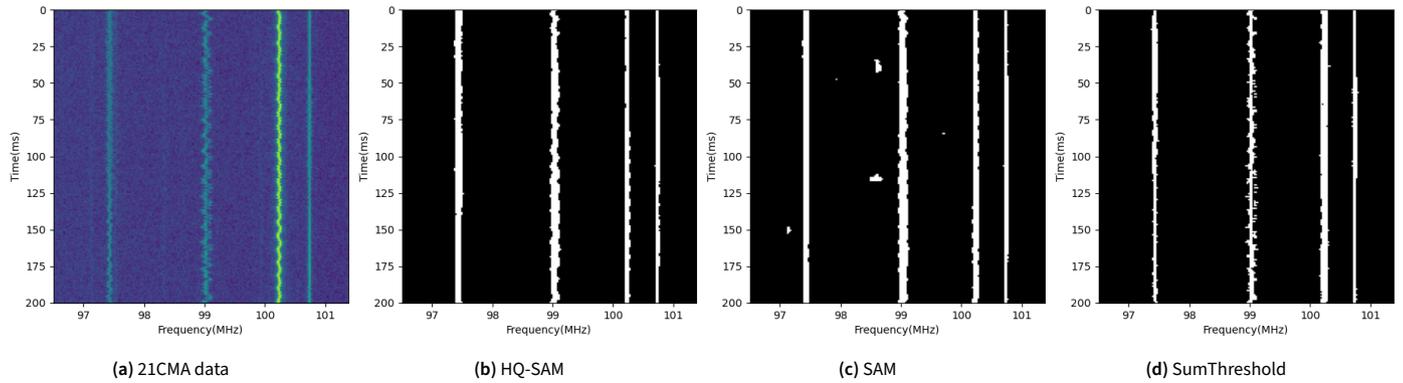

    \centering
    \begin{subfigure}{0.24\textwidth}
        \centering
        \includegraphics[width=1.15\linewidth]{fig/realdataresults/8_6_20000_96.5.pdf}
        \caption{21CMA data}
 
    \end{subfigure}
\hfill
    \begin{subfigure}{0.24\textwidth}
        \centering
        \includegraphics[width=1.15\linewidth]{fig/realdataresults/8_samhq6_20000_96.5.pdf}
        \caption{HQ-SAM}
      
    \end{subfigure}
 \hfill
    \begin{subfigure}{0.24\textwidth}
        \centering
        \includegraphics[width=1.15\linewidth]{fig/realdataresults/8_sam6_20000_96.5.pdf}
        \caption{SAM}
   
    \end{subfigure}
\hfill
    \begin{subfigure}{0.24\textwidth}
        \centering
        \includegraphics[width=1.15\linewidth]{fig/realdataresults/8_6_20000_96.5sum.pdf}
        \caption{SumThreshold}
     
    \end{subfigure}
    \caption{The raw data of the continuous narrowband RFI from the 21CMA and the detection results of three methods are shown in the following plots. The horizontal axis represents frequency, and the vertical axis represents time. (a) shows the raw data, which includes a large amount of continuous narrowband RFI. (b) shows the flagging results using HQ-SAM. (c) shows the flagging results using SAM. (d) shows the flagging results using the SumThreshold method. The white areas in the images correspond to the detected RFI.}
    \label{Fig2:continuous narrowband RFI}
\end{figure*}

\begin{figure*}[hbt!]
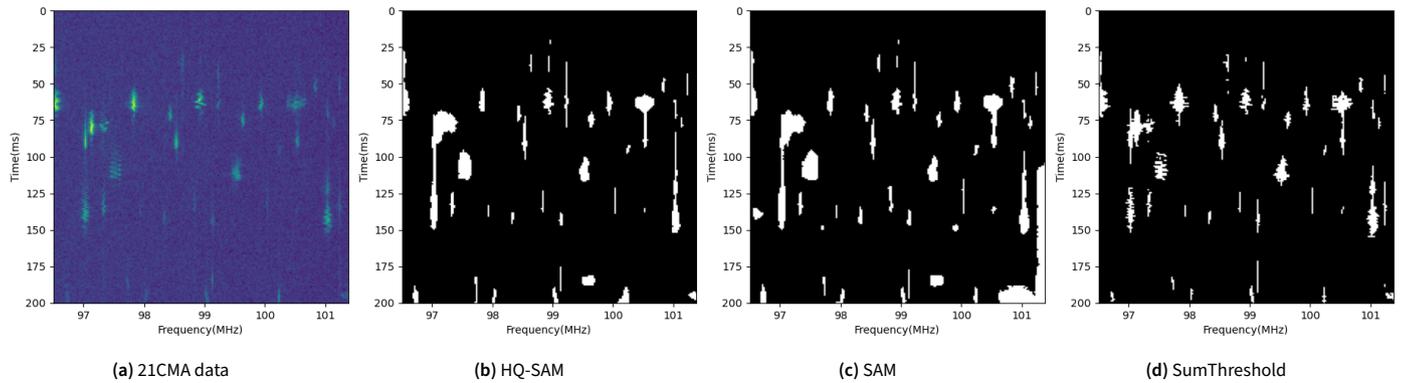

    \centering
    \begin{subfigure}{0.24\textwidth}
        \centering
        \includegraphics[width=1.15\linewidth]{fig/realdataresults/82_14600_96.5.pdf}
        \caption{21CMA data}
    
    \end{subfigure}
\hfill
    \begin{subfigure}{0.24\textwidth}
        \centering
        \includegraphics[width=1.15\linewidth]{fig/realdataresults/8samhq2_14600_96.5.pdf}
        \caption{HQ-SAM}
        
    \end{subfigure}
 \hfill
    \begin{subfigure}{0.24\textwidth}
        \centering
        \includegraphics[width=1.15\linewidth]{fig/realdataresults/8sam2_14600_96.5.pdf}
        \caption{SAM}
        
    \end{subfigure}
\hfill
    \begin{subfigure}{0.24\textwidth}
        \centering
        \includegraphics[width=1.15\linewidth]{fig/realdataresults/82_14600_96.5sum.pdf}
        \caption{SumThreshold}
      
    \end{subfigure}
    \caption{ The same as in Figure~\ref{Fig2:continuous narrowband RFI} but another example of the detection results of the 21CMA data.}
    \label{Fig3:transient narrowband RFI}
\end{figure*}

In the field of deep learning, it is common to compare detection results with a preset ground truth, providing quantitative metrics to reflect the model's capability. However, we consider the approach of flagging real data manually or using other deep learning methods as ground truth and comparing it with detection results to be not rigorous. This is because these labeling methods inevitably produce false positives and false negatives, which could affect the evaluation metrics. Therefore, in this part, we only qualitatively present the detection results, as shown in Fig.~\ref{Fig2:continuous narrowband RFI} and Fig.~\ref{Fig3:transient narrowband RFI}. The white areas in the images correspond to the detected RFI (the same for the subsequent images). For the quantitative comparison of the three methods, in Section~\ref{sect:RFI simulation}, we adopt an acceptable approach by simulating RFI. This allows us to artificially preset the RFI within the image as ground truth and compare it with the detection results to evaluate the model.

\subsection{The Results of HQ-SAM (SAM) of the 21CMA Data}
\label{subsec:classification}
\begin{figure*}[hbt!]
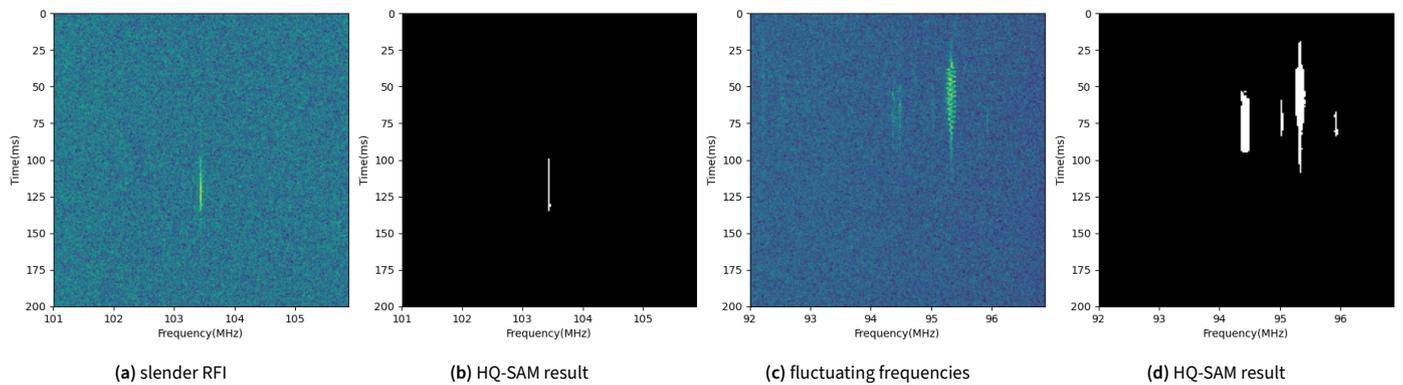

    \centering
    \begin{subfigure}[b]{0.24\textwidth}
        \centering
        \includegraphics[width=1.15\linewidth]{fig/RFI-classification/2_3800_101.0.pdf}
        \caption{slender RFI}
        \label{1}
    \end{subfigure}
    \hfill
    \begin{subfigure}[b]{0.24\textwidth}
        \centering
        \includegraphics[width=1.15\linewidth]{fig/RFI-classification/samhq2_3800_101.0.pdf}
        \caption{HQ-SAM result}
     
    \end{subfigure}
    \hfill
    \begin{subfigure}[b]{0.24\textwidth}
        \centering
        \includegraphics[width=1.15\linewidth]{fig/RFI-classification/3_17800_92.0.pdf}
        \caption{fluctuating frequencies}
        \label{fig:4.3}
    \end{subfigure}
    \hfill
    \begin{subfigure}[b]{0.24\textwidth}
        \centering
        \includegraphics[width=1.15\linewidth]{fig/RFI-classification/samhq3_17800_92.0.pdf}
        \caption{HQ-SAM result}
      
    \end{subfigure}
    \caption{Classification of narrowband RFI in frequency. (a) is a common slender signal. (b) is the result of detecting (a) using HQ-SAM. (c) is RFI with fluctuating frequencies, and (d) is the detection result of (c) by HQ-SAM.}
    \label{Frequency category}
\end{figure*}

\begin{figure*}[hbt!]
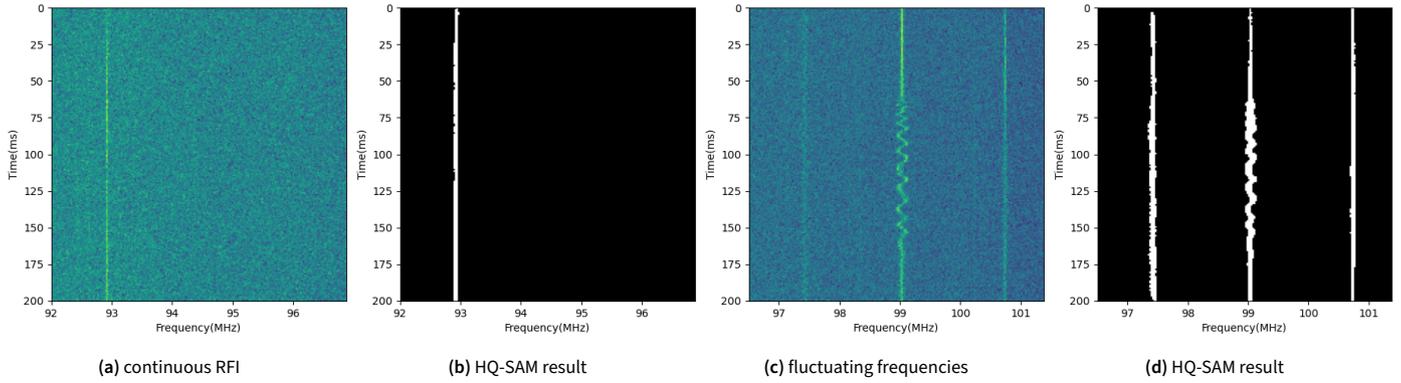

    \centering
    \begin{subfigure}[b]{0.24\textwidth}
        \centering
        \includegraphics[width=1.15\linewidth]{fig/RFI-classification/6_6400_92.0.pdf}
        \caption{continuous RFI}
       
    \end{subfigure}
    \hfill
    \begin{subfigure}[b]{0.24\textwidth}
        \centering
        \includegraphics[width=1.15\linewidth]{fig/RFI-classification/samhq6_6400_92.0.pdf}
        \caption{HQ-SAM result}
       
    \end{subfigure}
    \hfill
    \begin{subfigure}[b]{0.24\textwidth}
        \centering
        \includegraphics[width=1.15\linewidth]{fig/RFI-classification/6_5600_96.5.pdf}
        \caption{fluctuating frequencies}
      
    \end{subfigure}
    \hfill
    \begin{subfigure}[b]{0.24\textwidth}
        \centering
        \includegraphics[width=1.15\linewidth]{fig/RFI-classification/samhq6_5600_96.5.pdf}
        \caption{HQ-SAM result}
       
    \end{subfigure}
    \caption{Two types of continuous RFI. (a) and (c) are respectively slender RFI and RFI with fluctuating frequencies similar to frequency modulation. Unlike the RFI in Figure~\ref{Frequency category}, which last only around 50ms to 200ms, these RFI last for tens of seconds or more (the entire event of RFI is not fully displayed here).}
    \label{time category}
\end{figure*}

\begin{figure*}[hbt!]
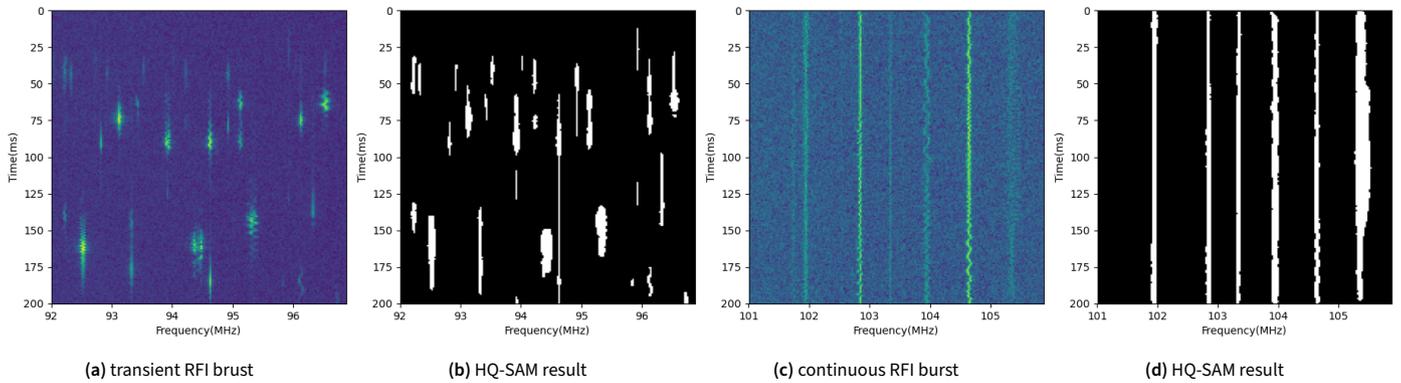

    \centering
    \begin{subfigure}[b]{0.24\textwidth}
        \centering
        \includegraphics[width=1.15\linewidth]{fig/RFI-classification/2_14600_92.0.pdf}
        \caption{transient RFI brust}

    \end{subfigure}
    \hfill
    \begin{subfigure}[b]{0.24\textwidth}
        \centering
        \includegraphics[width=1.15\linewidth]{fig/RFI-classification/samhq2_14600_92.0.pdf}
        \caption{HQ-SAM result}
        
    \end{subfigure}
    \hfill
    \begin{subfigure}[b]{0.24\textwidth}
        \centering
        \includegraphics[width=1.15\linewidth]{fig/RFI-classification/6_17000_101.0.pdf}
        \caption{continuous RFI burst}
        
    \end{subfigure}
    \hfill
    \begin{subfigure}[b]{0.24\textwidth}
        \centering
        \includegraphics[width=1.15\linewidth]{fig/RFI-classification/samhq6_17000_101.0.pdf}
        \caption{HQ-SAM result}
       
    \end{subfigure}
    \caption{The bursts of transient RFI and continuous RFI.}
    \label{quantity category}
\end{figure*}

\begin{figure*}[hbt!]
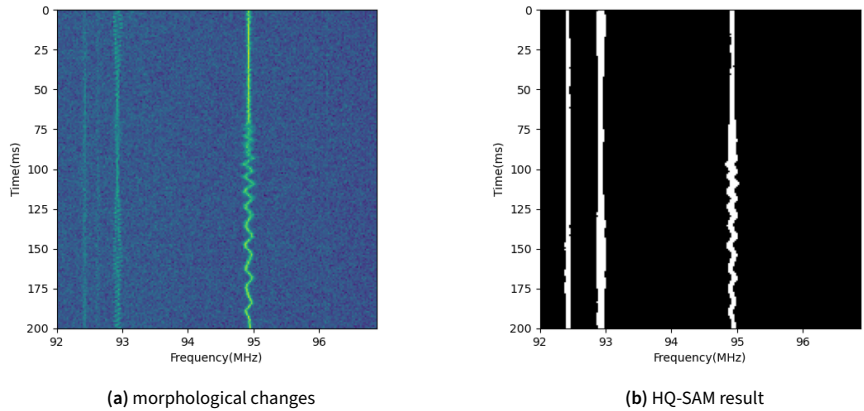

    \centering
    \begin{subfigure}[b]{0.3\textwidth}
        \centering
        \includegraphics[width=\linewidth]{fig/RFI-classification/6_13400_92.0.pdf}
        \caption{morphological changes }
       
    \end{subfigure}
    \hspace{6ex}
    \begin{subfigure}[b]{0.3\textwidth}
        \centering
        \includegraphics[width=\linewidth]{fig/RFI-classification/samhq6_13400_92.0.pdf}
        \caption{HQ-SAM result}
       
    \end{subfigure}

    \caption{transitions between straight lines and polylines during transmission for continuous RFI.}
    \label{polylines}
\end{figure*}

\begin{figure*}[hbt!]
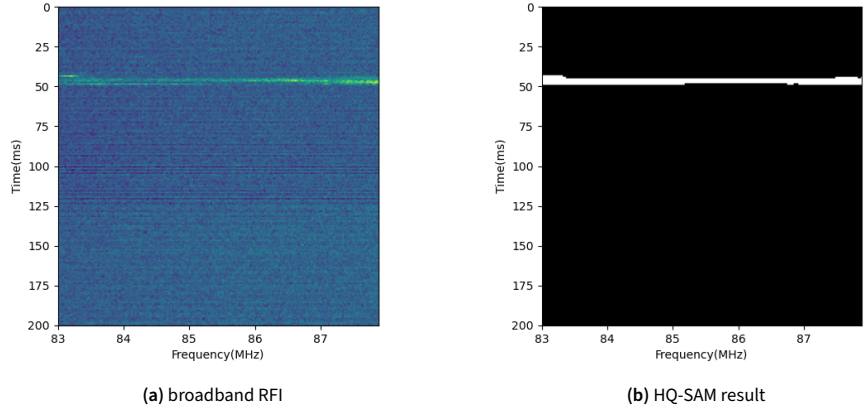

    \centering
    \begin{subfigure}[b]{0.3\textwidth}
        \centering
        \includegraphics[width=1\linewidth]{fig/RFI-classification/3_15200_83.0.pdf}
        \caption{broadband RFI}
      
    \end{subfigure}
    \hspace{6ex}
    \begin{subfigure}[b]{0.3\textwidth}
        \centering
        \includegraphics[width=1\linewidth]{fig/RFI-classification/samhq3_15200_83.0.pdf}
        \caption{HQ-SAM result}
      
    \end{subfigure}

    \caption{An example of broadband RFI which can contaminate frequency range of up to 20MHz.}
    \label{broadband}
\end{figure*}

We conducted a preliminary search of the 21CMA observational data (4.6 TB) using HQ-SAM and identified that the majority of the RFI consists of narrowband RFIs. Additionally, there are instances of broadband RFI events that contaminate a large number of frequency channels (see Fig.~\ref{broadband}). Based on the characteristics of RFI in time, frequency, and quantity, we divide narrowband RFI into three main categories, while broadband RFI is treated separately as a special case. As shown in Fig.~\ref{Frequency category}, in addition to the common slender signals, narrowband RFIs demonstrate a frequency modulation pattern similar to FM radio signals. Fig.\ref{time category} shows that the duration of narrowband RFI can mainly be classified into two groups. Apart from the transient or short-duration signals with durations around 50 ms to 200 ms shown in Fig.\ref{Frequency category}, there are numerous continuous RFIs lasting for tens of seconds or more (the entire event of RFI is not fully displayed in Fig.~\ref{time category}). Furthermore, narrowband RFI is observed both sporadically as depicted in Fig.~\ref{Frequency category} and~\ref{time category}, and in large bursts dispersed across multiple frequency bands, as illustrated in Fig.~\ref{quantity category}. For continuous narrowband RFI, it is intriguing to find transitions between straight lines and polylines during transmission (Fig.\ref{polylines}). Fig.\ref{broadband} illustrates the broadband signals detected by us, which span a frequency range of up to 20 MHz.

According to \citet{gao}, we believe that the primary cause of transient narrowband RFI is the scattering of FM broadcasts by meteor trails. For the continuous narrowband signals, the number of events varies with time, similar to the variation in meteor events due to the Earth's rotation. The quantity of narrowband RFIs far exceeds that of the known nearby FM broadcasts. In fact, there are no constant FM radio signals at the 21CMA site. Apart from the narrowband signals originating from FM broadcasts, the remaining sources require further investigation.

In summary, we find that HQ-SAM (SAM) performs comparably to the SumThreshold method when applied to real observational data from the 21CMA, especially in finding as many RFIs as possible. We conducted a sanity check of the detection results by comparing the results from HQ-SAM (SAM) with those from human visual inspection. HQ-SAM (SAM) effectively identifies most of the RFIs detected by human visual inspection. Furthermore, HQ-SAM outputs less noise and higher quality masks compared to SAM in actual detection results, thus confirming the assertion by \citet{hqsam}.In Appendix~\ref{Real data appendix}, the results of the three methods for Fig.~\ref{Fig2:continuous narrowband RFI} and Fig.~\ref{Fig3:transient narrowband RFI} are plotted pairwise on the same layer to show the differences in the masks obtained by each method.

However, HQ-SAM (SAM) also presents several issues worthy of optimization: when two RFI events are too close in the waterfall plot, the model may classify them as a single RFI event; there is a certain amount of noise present in the output results; the segmentation of RFI profiles is relatively coarse; and although HQ-SAM (SAM) can identify some faint RFI, it still has limitations with extremely faint RFI signals, which are barely noticeable to the human eye as well. The SumThreshold method faces similar challenges.

\section{RFI Detection of Simulation Data  }
\label{sect:RFI simulation}

In this section, by mock RFI data, we can further evaluate the performance of the three methods and provide specific evaluation metrics to illustrate their respective strengths and weaknesses.

\subsection{RFI Simulation}

\begin{figure}[hbt!]
   \centering
   \includegraphics[width=1\columnwidth, angle=0]{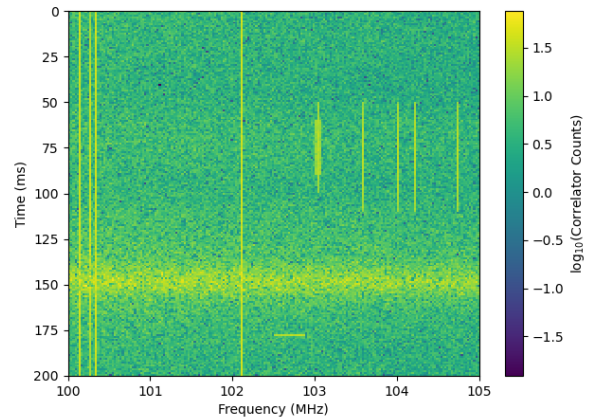}
   \caption{An example of Type A mock RFI, which contains dtv RFI and narrowband RFI.}
   \label{mockrfi1}
   \end{figure}

We use the hera\_sim\footnote{https://github.com/HERA-Team/hera\_sim} to simulate RFI signals. The hera\_sim is a basic simulation package for HERA-like redundant interferometric arrays, which can also generate RFI (\citealt{chen2015using, Kerrigan_2019, cnn-sun, liang2023detecting}).
There are two types of RFI data that have been created, each consisting of 300 waterfall plots with RFI signals, all of which are 200$\times$200 in size. The horizontal axis of these waterfall plots represents frequency, and the vertical axis represents time. For the first type of mock RFI (as shown in Fig.\ref{mockrfi1}), referred to as Type A, basic thermal noise and EoR-like visibilities are simulated using hera\_sim. This tool allows us to generate several types of RFI, and we select narrowband RFI and RFI arising from digital TV channels (DTV RFI). The DTV RFI are set to be distributed with a probability of 0.005 over a certain frequency range and appear as rectangular blocks, with an amplitude approximately 0.71 magnitudes higher than the background (in base-10 logarithm; the following is simplified to approximate numbers). Considering the characteristics mentioned in Section\ref{sect: real RFI}, including intensity variations, frequency fluctuations, and differences in durations (transient or continuous), we adjust the code for narrowband RFI to better reflect real-world scenarios. Hence, the following sorts of narrowband RFI can be observed in Fig.~\ref{mockrfi1}: continuous RFI with fixed intensity (0.84), continuous and transient RFI with intensity varying over time (0.54-0.84; 0.54-0.71), and a transient event with a noticeable frequency width (a simple imitation of RFI in Fig.~\ref{fig:4.3} which has frequency fluctuations; 0.54-0.71). Within their respective frequency ranges, the probabilities of these events are 0.01, 0.01, and 0.04, respectively, and the transient event is set to appear with a probability of 0.6 per picture. It is worth mentioning that the SEEK package performs median normalization on the data before applying the SumThreshold method.  If there are continuous narrowband RFI that do not vary dramatically in intensity, they will be removed at this step, resulting in their undetection.  We speculate that this may be because the designers did not anticipate facing this kind of RFI mitigation. In this paper, the code is modified to remove this normalization step.

\begin{figure}[hbt!]
   \centering
   \includegraphics[width=1\columnwidth, angle=0]{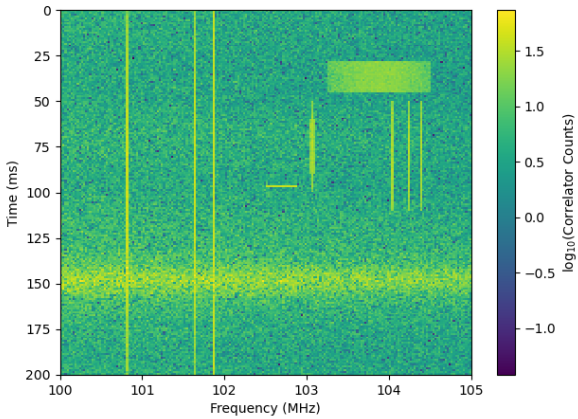}
   \caption{ An example of Type B mock RFI. Compared to Type A, it includes an additional type of broadband RFI.}
   \label{mockrfi2}
   \end{figure}

Inspired by the broadband streaks reported in \citet{Wilensky_2019} and the broadband event shown in Fig.~\ref{broadband}, the second type of mock RFI, Type B, introduces a new category of broadband signals (0.24-0.54) with a probability of 0.6 per picture, while the other aspects remain the same as in the first type. Fig.~\ref{mockrfi2} is an example of such simulated signals. There is a broadband event with intensity changing by frequency, set as a rectangle.

\subsection{Evaluation Metrics}
\label{evaluation}

By simulating, we obtain spectra data containing the RFI and the corresponding ground truth. Then, utilize HQ-SAM, SAM, and the SumThreshold method to recognize the mock signals and, respectively, compare the recognition results with the ground truth to get quantitative evaluation metrics.

The evaluation metrics used in this article include precision, recall, precision, and F1 score. They are defined as follows:
\begin{equation}\label{eq3}
    {\rm Accuracy} = \frac{\rm TP+TN}{\rm TP+TN+FP+FN}
\end{equation}
\begin{equation}\label{eq4}
    {\rm Recall} = \frac{\rm TP}{\rm TP+FN}
\end{equation}
\begin{equation}\label{eq5}
    {\rm Precision} = \frac{\rm TP}{\rm TP+FP}
\end{equation}
\begin{equation}\label{eq6}
    {\rm F1} = \frac{2\times{\rm Precision}\times{\rm Recall}}{\rm Precision+Recall}
\end{equation}
In this work, the counts of true positive (TP), true negative (TN), false positive (FP), and false negative (FN) are based on individual pixels, where each pixel is classified as either RFI or non-RFI.

The evaluation results of the three methods for the Type A recognition task are shown in Table~\ref{Tab1}. And the evaluation results for Type B are shown in Table~\ref{Tab2}.

\begin{table}[hbt!]

\caption{Accuracy, Recall, Precision, and F1 Score of three methods for the Type A mock RFI recognition task.}
\label{Tab1}
\begin{tabular}{ccccc}
\toprule
\headrow  & Accuracy   & Recall & Precision & F1 Score \\
\midrule
HQ-SAM  & 99.67\% & 97.74\%     & 87.67\% &  92.43\%          \\
\midrule
SAM  &99.55\%   &   97.12\%     & 82.04\%     & 88.95\%           \\
\midrule
SumThreshold & 99.97\% &  98.38\%   & 98.98\%  &98.68\%   \\
\bottomrule
\end{tabular}

\end{table}

\begin{table}[hbt!]
\caption{ Accuracy, Recall, Precision, and F1 Score of three methods for the Type B mock RFI recognition task.}
\label{Tab2}
\begin{tabular}{ccccc}
\toprule
\headrow  & Accuracy   & Recall & Precision & F1 Score \\
\midrule
HQ-SAM  & 99.66\% & 98.19\%     & 91.54\% &  94.75\%          \\
\midrule
SAM  & 99.29\%   & 97.27\%     & 82.35\%     & 89.19\%           \\
\midrule
SumThreshold  & 98.90\%     &   74.66\%   & 93.80\%  &83.14\%                \\
\bottomrule
\end{tabular}
\end{table}

\subsection{Results}

According to Section~\ref{evaluation}, we are able to provide quantitative evaluations for the three methods.

In RFI recognition tasks, RFI typically exhibit complex structures and characteristics. In the two types of simulated data, HQ-SAM outperforms SAM in terms of metrics (especially Precision), indicating that HQ-SAM demonstrates better performance than SAM, thus confirming the claims made in \citet{hqsam}.

For Type A mock data recognition, the SumThreshold method shows excellent performance with an F1 Score of 98.68\%. HQ-SAM, without fine-tuning or structural additions, also performs commendably with an F1 Score of 92.43\%. SAM, while capable of recognizing most RFI, shows worse performance in Precision compared to the first two methods. This is because, under the same conditions, SAM outputs more miscellaneous items than HQ-SAM, which lowers the Precision score and, consequently, the F1 Score.

For Type B mock data recognition, where broadband RFI or large areas of RFI are present, HQ-SAM (and SAM) demonstrates superior performance compared to the SumThreshold method, with an F1 Score of 94.75\% compared to 83.14\%. Notably, HQ-SAM (and SAM) excels in Recall, achieving 98.19\% versus 74.66\% for the SumThreshold method. This high Recall is particularly crucial for our goal of identifying and removing as many RFI events as possible.

Since SAM outputs multiple masks and we use automatic evenly sprinkling points as prompts, there will be miscellaneous masks mixed with the masks we need. We use mask area and predicted Intersection over Union (IoU) limits to filter out these miscellaneous masks. Therefore, if one plans to use it, additional filtering conditions may be necessary from the start. In fact, manual inspection reveals that SAM and HQ-SAM actually achieve better performance metrics than those listed. This discrepancy arises because some miscellaneous and over-recognition masks, which overwrite the precise segmentation masks, were not filtered by our simple conditions, leading to a lower Precision score than what could actually be achieved.

In addition to the problems mentioned in Section~\ref{subsec:classification}, we note that for continuous narrowband RFI, which morphologically presents as a straight line and lasts the entire observation time, if it is too close to the edge of the image, the model may recognize the region between it and the boundary without properly identifying the narrowband itself. Similarly, when two narrowbands are too close together, both the RFI and the region between them may be identified as a single entity. As a result, the output masks in these cases are filtered out, which reduces the Recall score.

In summary, we conlcude that, on one hand, HQ-SAM can be used directly to identify RFI due to its strong generalization capability and recognition ability. On the other hand, it is worth fine-tuning HQ-SAM (SAM) in this domain to optimize performance, or developing better automatic mask filtering methods to achieve more precise segmentation results.

\subsection{Resource Requirements and Speed}

Compared to the SumThreshold method, HQ-SAM (and SAM) does not require manual adjustment of thresholds and iteration counts. However, it does incur additional GPU requirements (the model can run on both CPU and GPU). Our server is equipped with a Tesla P100 PCIe 16GB GPU. SAM using the vit\_h checkpoint requires 6.5GB of GPU memory, while HQ-SAM requires 10.5GB.

Besides calculating evaluation metrics, we also measure the runtime of the three methods. HQ-SAM (and SAM) require more time compared to the SumThreshold method. Under the same conditions for HQ-SAM and SAM ($\rm{points\_per\_side = 96}$), SAM takes approximately four times longer than the SumThreshold method, while HQ-SAM takes nearly six times longer.

For SAM, when $\rm{points\_per\_side = 40}$, the runtime is comparable to that of the SumThreshold method. For HQ-SAM, however, $\rm{points\_per\_side = 34}$ is required. In this case, for both Type A and Type B mock data recognition, Recall of the two models decrease significantly (for Type A, Recall of SAM and HQ-SAM decrease by 33\% and 45\%, respectively, while for Type B, the decreases are 20\% and 30\%), while Precision change little. Therefore, we do not recommend excessively reducing computational performance for the sake of speed. Regarding Type B, if we compare the performance of the three method using F1 Score, SAM and HQ-SAM achieve F1 Score that are roughly comparable to the SumThreshold method when $\rm{points\_per\_side = 64}$ and $\rm{points\_per\_side = 40}$, respectively, with their runtime being approximately 2.2 times and 1.3 times that of the SumThreshold method. Specific details about the evaluation metrics can be found in Appendix~\ref{appendix speed}.

In some scenarios that require real-time processing, SAM may not be suitable due to the significant computational cost introduced by its image encoder. Many works have attempted to improve the efficiency of SAM. For example, MobileSAM (\citealt{MobileSAM}) distills the knowledge from the heavy image encoder to a lightweight one. To enhance the speed of SAM in recognizing RFI, replacing the image encoder with a lightweight model may be a promising direction for our future efforts.

\section{Solar Radio Burst Detection by HQ-SAM}
\label{sect:SRB}

The intricate characteristics of solar radio bursts pose significant challenges for their automatic detection and classification. Currently, there are few studies on deep learning for SRB detection, and they are often limited by insufficient training data. Many of these studies adopt methods such as transfer learning \citep{SRB} or SRB simulation \citep{Scully_2023SRB} to address this problem. Here, we apply HQ-SAM in the detection of SRBs, aiming to determine whether it demonstrates strong zero-shot and generalization capabilities in such applications.

\subsection{Solar Radio Burst Data}

\begin{figure*}[hbt!]
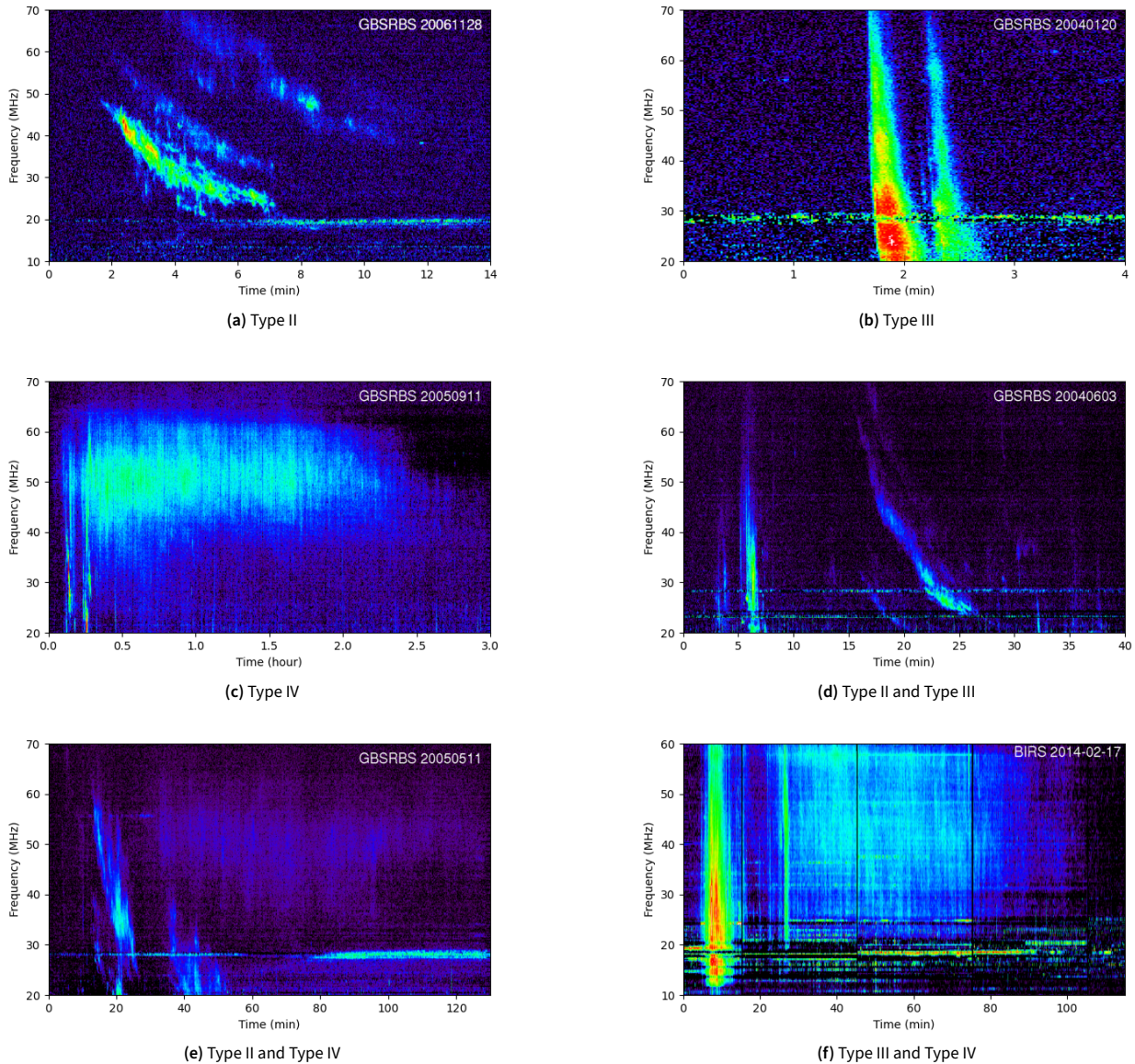

    \centering
    \begin{subfigure}{0.45\textwidth}
        \centering
        \includegraphics[width=1\linewidth]{fig/SRB/SRB-examples/II20061128.pdf}
        \caption{Type II}
   
    \end{subfigure}
    \hspace{6ex}
    \vspace{1ex}
    \begin{subfigure}{0.45\textwidth}
        \centering
        \includegraphics[width=1\linewidth]{fig/SRB/SRB-examples/III20040120.pdf}
        \caption{Type III}
    
    \end{subfigure}
 
    \begin{subfigure}{0.45\textwidth}
        \centering
        \includegraphics[width=1\linewidth]{fig/SRB/SRB-examples/IV20050911.pdf}
        \caption{Type IV}
      
    \end{subfigure}
   \hspace{6ex}
    \begin{subfigure}{0.45\textwidth}
        \centering
        \includegraphics[width=1\linewidth]{fig/SRB/SRB-examples/II_III20040603.pdf}
        \caption{Type II and Type III}
       
    \end{subfigure}

    \begin{subfigure}{0.45\textwidth}
        \centering
        \includegraphics[width=1\linewidth]{fig/SRB/SRB-examples/II_IV20050511.pdf}
        \caption{Type II and Type IV}
      
    \end{subfigure}
   \hspace{6ex}
    \begin{subfigure}{0.45\textwidth}
        \centering
        \includegraphics[width=1\linewidth]{fig/SRB/SRB-examples/III_IV20140217.pdf}
        \caption{Type III and Type IV}
      
    \end{subfigure}

    \caption{Several types of SRB data used in the Detection. There are Type II-IV alone and in pairs. The horizontal axis represents time, while the vertical axis represents frequency.}
    \label{SRB data example}
\end{figure*}

The SRB data for this study are images of SRBs observed by the Green Bank Solar Radio Burst Spectrometer (GBSRBS), provided by the National Radio Astronomy Observatory website\footnote{https://www.astro.umd.edu/~white/gb/}. The horizontal axis represents time, while the vertical axis represents frequency. According to the shape, frequency, and time length of the SRBs, these images are classified into three types: Type II, Type III, and Type IV \citep{SRB}. We obtained a total of 636 images, all cropped to the size of 1225$\times$645. Fig.~\ref{SRB data example} shows several types of SRB data: Type II-IV alone and in pairs.

\subsection{Detection Results by HQ-SAM}
\label{HQ-SAM SRB}

\begin{figure*}[hbt!]
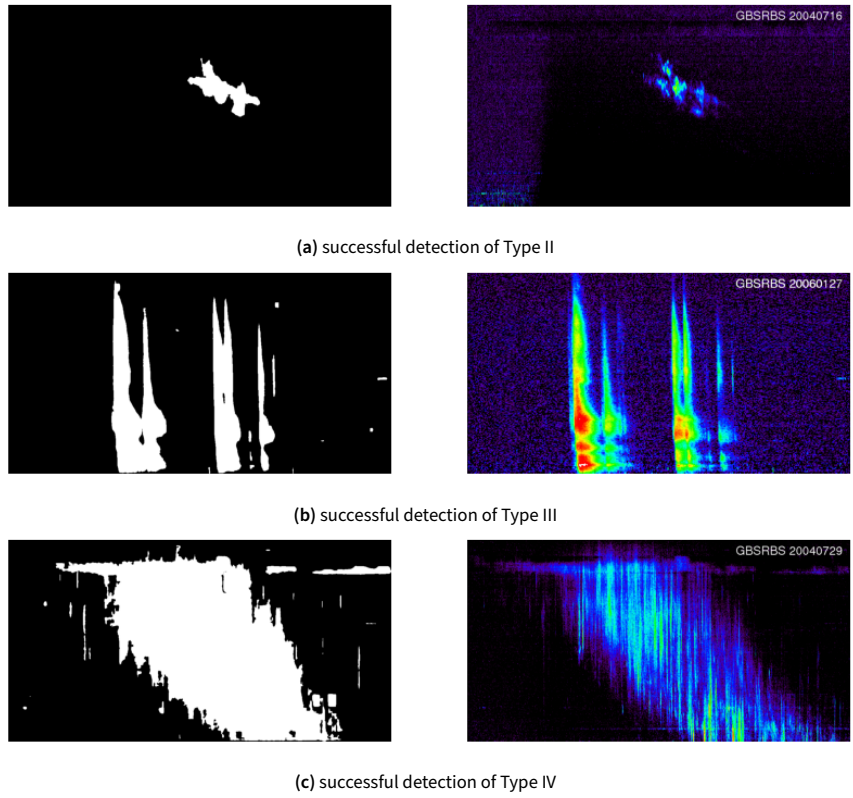

    \centering
    \begin{subfigure}{0.7\textwidth}
        \centering
        \includegraphics[width=0.9\linewidth]{fig/SRB/SRBresults/success/typeII_1.pdf}
        \caption{successful detection of Type II}       
    \end{subfigure} 
    \begin{subfigure}{0.7\textwidth}
        \centering
        \includegraphics[width=0.9\linewidth]{fig/SRB/SRBresults/success/typeIII_1.pdf}
        \caption{successful detection of Type III}
   
    \end{subfigure}
  
    \begin{subfigure}{0.7\textwidth}
        \centering
        \includegraphics[width=0.9\linewidth]{fig/SRB/SRBresults/success/typeIV_1.pdf}
        \caption{successful detection of Type IV}
     
    \end{subfigure}    
    \caption{Examples of successful detection for Type II-IV SRBs. The right side is the raw data, and the left side is the output masks.}
    \label{success}
\end{figure*}

\begin{figure*}[hbt!]
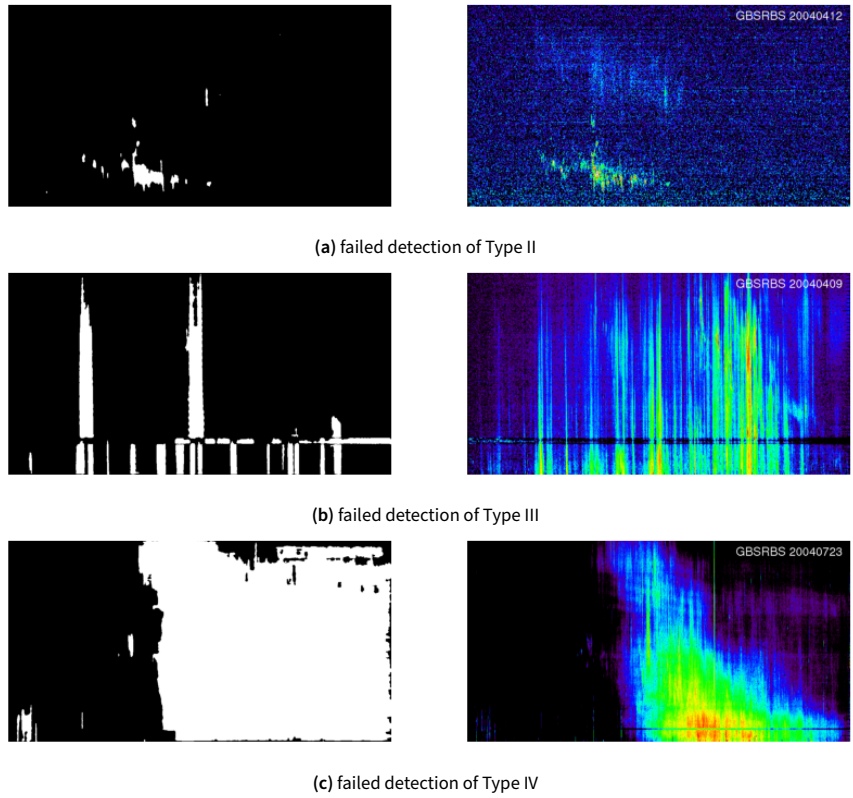

    \centering
    \begin{subfigure}{0.7\textwidth}
        \centering
        \includegraphics[width=0.9\linewidth]{fig/SRB/SRBresults/failure/failII_20040409_200500_BIRS.pdf}
        \caption{failed detection of Type II}
     
    \end{subfigure} 
    \begin{subfigure}{0.7\textwidth}
        \centering
        \includegraphics[width=0.9\linewidth]{fig/SRB/SRBresults/failure/failIII_20040119_200500_BIRS.pdf}
        \caption{failed detection of Type III}
       
    \end{subfigure}

    \begin{subfigure}{0.7\textwidth}
        \centering
        \includegraphics[width=0.9\linewidth]{fig/SRB/SRBresults/failure/failIV_20040723_181500.pdf}
        \caption{failed detection of Type IV}
        
    \end{subfigure}    
    \caption{Examples of detection failure for Type II-IV SRBs.}
    \label{failure}
\end{figure*}

Here, we present the detection results of HQ-SAM for SRB data in the form of output masks. Our goal is to recognize the presence of all SRB events in the data and accurately identify the main portions of these events. A small amount of over-marking or under-marking is acceptable. We overlay the masks onto the original images and manually check for differences. Detection is considered successful if the events are recognized and the corresponding masks align well with the SRB contours in the image. If events are undetected, detected incorrectly, or if the masks over-label or under-label by more than 30\% area, even if the presence of the event has been detected, it is deemed a detection failure. Fig.\ref{success} shows examples of successful detection for separate Type II-IV, while Fig.\ref{failure} demonstrates examples of failures for the corresponding types. A more comprehensive display of the detection results for all six types can be found in Appendix~\ref{SRB detection results appendix}. In the qualitative presentation of these results, the axes are omitted.

Table~\ref{Tab3} shows the ratio of successful detections by HQ-SAM in recognizing various SRBs, determined by manual inspection.

\begin{table}[hbt!]

\caption{ The ratio of successful detection by HQ-SAM in recognizing various SRBs.}
\label{Tab3}
\begin{tabular}{ccc}
\toprule
\headrow Type & Total quantity& Ratio of successful detection             \\
\midrule
Type II  & 108 & 87.04\%         \\
\midrule
Type III  & 301   &   96.68\%         \\
\midrule
Type IV  & 78    &  44.87\%             \\
\midrule
Type II/III  & 87    &  90.80\%             \\
\midrule
Type II/IV  & 44  &  45.45\%             \\
\midrule
Type III/IV  & 18    &  22.22\%             \\
\bottomrule
\end{tabular}
\end{table}

\subsection{Results}

When appearing isolated in spectrograms i(e.g., Fig.\ref{success} and Fig.\ref{failure}), Type II and Type III SRBs can be effectively identified. However, HQ-SAM's ability to recognize Type IV SRBs is weaker than for the first two types. Despite this, nearly half of the Type IV instances in this study are successfully detected.

In cases where two types of SRB events coexist (e.g., Fig.\ref{success supplement2} and Fig.\ref{failure supplement}), effective detection is achieved when Type II and Type III occur simultaneously. However, when Type IV coexists with the other two types, recognizing the Type IV is challenging. If there is overlap between Type IV and any of the other types, it further interferes with the detection of Type II and Type III because of Type IV morphological dispersion. During manual inspection, we notice that failures in detecting Type II/IV or Type III/IV data often occur due to difficulties in identifying Type IV SRBs, while Type II and Type III are frequently successfully identified.

Particularly, the large area, relatively diffuse distribution, and low intensity of Type IV SRBs—which often appear dim on spectrograms—make it challenging for the model to capture their features, affecting detection accuracy. Even when HQ-SAM successfully identifies Type IV SRBs, there may be errors in determining the exact contours of these events, leading to issues of over-labeling or under-labeling. For instance, Type IV SRBs and their surrounding background may sometimes be recognized as a single entity, resulting in masks that over-label with large areas. This issue is illustrated in Fig.\ref{large area mask1} and Fig.\ref{large area mask2}, and such masks are often filtered out. The rare cases of detection failures of Type II, Type III, and Type II/III suggest that when these types of SRB or their backgrounds are too dim or diffuse, under-labeling can also occur.

In addition, the model's inherent characteristics, instrument-generated stripes, and noise in the spectrogram can introduce various artifacts into the output masks, including small random points and large erroneous areas. Without appropriate filtering criteria, even if HQ-SAM accurately identifies the main structures of SRBs, the results may still be contaminated by these artifacts. Therefore, establishing effective filtering conditions for the masks is crucial.

Overall, HQ-SAM exhibits impressive performance in SRB detection tasks, particularly in identifying Type II and Type III SRBs. This is notable given that these results are achieved without additional model optimization, and the model and its weights are readily accessible from the relevant website. Thus, we believe it is highly worthwhile to further explore the application of HQ-SAM in the field of SRB recognition.

\section{Discussion}
\label{disccussion}
We have demonstrated that SAM and HQ-SAM hold significant potential for detecting various types of RFI and events. However, there is still room for improvement in the SAM and HQ-SAM models. For generalized models, specifying prompts helps filter the output masks, ensuring that the results focus on areas of interest. Since SAM cannot automatically generate suitable prompts for itself, we use the automatic segmentation mode provided by SAM, which involves evenly distributing points across the image. This method may lead to insufficient attention to important parts of the image. Alternatively, SAM offers the option to manually provide point or box prompts. Our experiments show that manually providing points results in more accurate target recognition compared to the automatic mode.

\citet{cropsambased} suggests using YOLO-v8 for initial image detection and obtaining bounding boxes around targets, which then serve as prompts to enhance segmentation quality. While this approach can improve precision in segmentation tasks, introducing a trained YOLO-v8 might reduce the model's ability to detect unknown phenomena when aiming to identify all potential targets. A possible solution is to combine evenly distributed points with YOLO-v8, assigning different weights to each method based on specific needs. This combination could potentially improve recognition performance while preserving the model's capacity to identify unknown events.

Additionally, implementing appropriate transfer learning for the model could be advantageous. Analysis of detection results from real and simulated RFI provided by HQ-SAM reveals that the model's prior training, which primarily involved rich landscape images, inadequately addressed dim line images. This discrepancy sometimes leads the model to mistakenly interpret an RFI line as a boundary of an adjacent area. Therefore, targeted transfer learning can enhance the model's ability to recognize and understand dim line features without introducing excessive data dependency or diminishing its capability to detect unknown phenomena.

It has been observed that the color of an image significantly impacts the model's recognition performance.  The HQ-SAM model shows varying performance with different color schemes. Specifically, for the simulated narrowband RFI discussed in Section~\ref{sect:RFI simulation}, converting RGB images to pseudocolor grayscale  produces coarser output masks, causing Precision to decline significantly (generally over 30\%), as shown in Table~\ref{color table1} and Table~\ref{color table2}.

These variations are likely due to the high-dimensional nature of RGB values, where even minor color changes can produce significantly different feature vectors. This variability affects the model's recognition accuracy. Therefore, it is crucial to consider factors such as image color, detection target size, and image resolution when setting recognition targets to ensure optimal performance of the model.

\section{Conclusions}
\label{sect:conclusion}

In this paper, we apply HQ-SAM (SAM) to various scenarios of RFI and event detection in radio astronomy, demonstrating its impressive recognition and generalization capabilities. For RFI detection, HQ-SAM (SAM) is utilized to identify both real RFI data from the 21CMA and simulated RFI data generated with the hera\_sim package. The performance of HQ-SAM is compared with the SumThreshold method, showing superior results. Specifically, for broadband or large-area mock RFI, HQ-SAM achieves higher recall compared to the SumThreshold method, effectively reducing false negatives. Additionally, HQ-SAM demonstrates strong recognition abilities in detecting solar radio bursts, including Type II and Type III bursts.

However, the model has some limitations, including the need for additional filtering due to miscellaneous items in the recognition results, coarse mask profiles, the computational limitations caused by the heavy image encoder, and the absence of semantic categorization. Deep learning models often face challenges such as insufficient training data, data imbalance, over-reliance on training sets, and limited generalization capabilities, as they are typically designed for specific tasks. Furthermore, designing and training models from scratch is a significant burden. HQ-SAM (SAM) stands out with its powerful generalization capabilities, enabling easy deployment without extensive transfer learning or modifications. It functions as a plug-and-play solution and may help address or mitigate these common issues, making it a promising candidate for broader applications in astronomy.

\begin{acknowledgement}
We acknowledge the support of the Ministry of Science and Technology of China (grant No. 2020SKA0110200). YY acknowledges the support of the Key Program of National Natural Science Foundation of China (12433012).
\end{acknowledgement}

\paragraph{Competing Interests}

None

\paragraph{Data Availability Statement}

The data and code underlying this article can be shared on reasonable request to the corresponding author.

\printendnotes

\printbibliography

\appendix

\section{Supplement to the SumThreshold Method}
\label{sumthreshold appendix}
For two neighboring samples, $A$ and $B$, traditional thresholding involves independently comparing a statistic from each sample to a fixed threshold, which can lead to false positives and false negatives. In contrast, the VarThreshold method improves upon this by using combinatorial thresholding. It iteratively combines samples and compares them against a strictly decreasing series of thresholds, $\left \{ \chi _i \right \} ^N_{i=1}$, where $N$ is the number of iterations.

Initially, if $A$ and $B$ individually do not exceed the first threshold, $\chi_1$, they are evaluated together with a lower threshold, $\chi_2$. If both samples exceed $\chi_2$, they are flagged. Otherwise, $A$, $B$, and the next neighboring sample, $C$, are combined and compared with a further threshold, $\chi_3$ \citep{Offringa_2010}. This process continues iteratively and can be represented as:

\begin{equation}\label{eq1}
\begin{aligned}
     &flagv_M(v,t)=\exists i\in \left \{ 0...M-1 \right \} : \\
     &\forall j\in \left \{0...M-1  \right \} :\left | R(v+(i-j)\Delta_v,t) \right |>\chi_{N(M)}
\end{aligned}
\end{equation}
where $M$ is the number of samples. If the absolute values of all samples exceed the threshold $\chi_{N(M)}$, these samples are flagged as RFI. Empirically, $M = \left \{ 1, 2, 4, 8, 16, 32, 64 \right \} $ has been found to be most effective and time-efficient, corresponding to 7 iterations.

The strictly decreasing series of thresholds, $\left \{ \chi_i \right \}_{i=1}^N$, is given by:
\begin{equation}\label{eq2}
    \chi_i = \frac{\chi_1}{\rho^{\log_{2}{i} }}
\end{equation}
Empirically, $\rho = 1.5$ is a suitable choice for both the VarThreshold and the subsequent SumThreshold methods.

The SumThreshold method calculates the sum of statistics from $M$ samples and compares it to $M$ times the corresponding threshold, $\chi_{N(M)}$. If samples flagged as RFI in previous iterations are encountered again, their values are replaced by the current iteration's threshold. For example, consider the sample set $\left [ 2, 2, 5, 7, 2, 2 \right ]$, where $\left [5,7\right ]$ are the RFI to be flagged. With thresholds $\left (6,4,3.16 \right )$ for 3 iterations, and sample numbers $\left (1,2,4 \right )$, without applying the condition, $\left [ 2,2,5,7,2,2 \right ]$ would yield 4 false positives. By applying the condition, $\left [ 7 \right ]$ is flagged first, resulting in samples $\left [ 2,2,5,4,2,2 \right ]$. Then, $\left [ 5,4 \right ]$ are flagged, and in the third iteration, $\left [ 2,2,3.16,3.16,2,2 \right ]$ yields no additional flags. Thus, only $\left [5,7 \right ]$ are flagged, avoiding false positives.

The SumThreshold method also allows for flagging sequences of samples with values below the thresholds. For a sample set $\left [ 1,3,4,7,4,3,1 \right ]$, where $\left [ 3,4,7,4,3 \right ]$ represent RFI with varying intensities, setting thresholds to $\left [ 6,4,3.16 \right ]$ and sample numbers to $\left [ 1,2,4 \right ]$, the VarThreshold method flags only $\left [ 7 \right ]$, whereas the SumThreshold method flags $\left [ 3,4,7,4,3 \right ]$.

\section{Supplement to RFI Detection Results of Real Data}
\label{Real data appendix}

\begin{figure*}[hbt!]
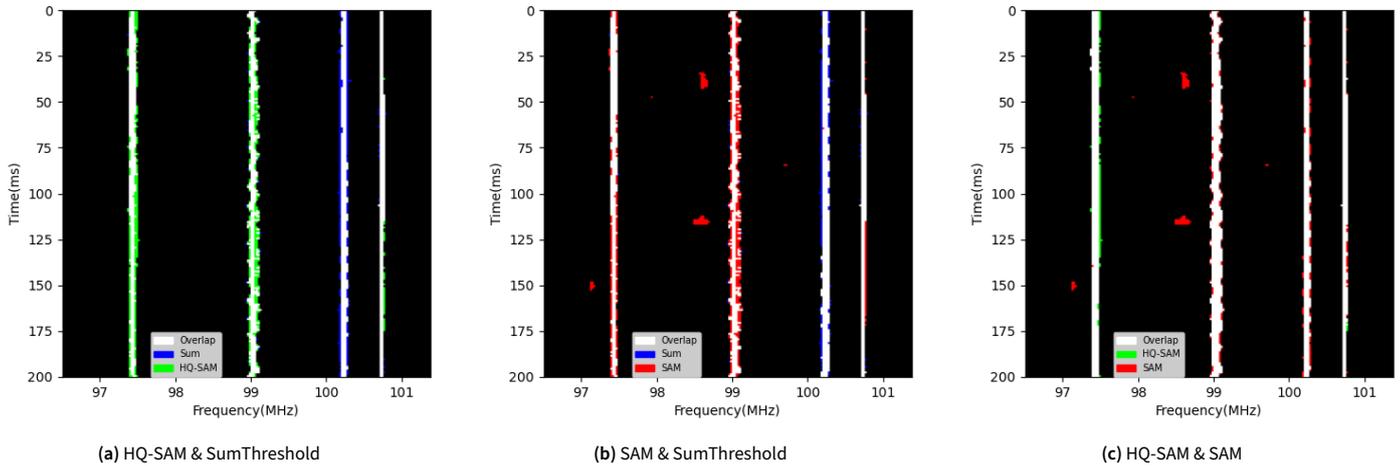

    \centering

    \begin{subfigure}{0.3\textwidth}
        \centering
        \includegraphics[width=1.15\linewidth]{fig/RFIresultcomparison/samhq-sum-6_20000.pdf}
        \caption{HQ-SAM \& SumThreshold}
       \label{comparison-2-b}
    \end{subfigure}
 \hfill
    \begin{subfigure}{0.3\textwidth}
        \centering
        \includegraphics[width=1.15\linewidth]{fig/RFIresultcomparison/sam-sum-6_20000.pdf}
        \caption{SAM \& SumThreshold}
   \label{comparison-2-c}
    \end{subfigure}
\hfill
    \begin{subfigure}{0.3\textwidth}
        \centering
        \includegraphics[width=1.15\linewidth]{fig/RFIresultcomparison/sam-samhq-6_20000.pdf}
        \caption{HQ-SAM \& SAM}
     \label{comparison-2-d}
    \end{subfigure}
    \caption{
    The differences in the masks obtained by each methods for Fig.~\ref{Fig2:continuous narrowband RFI} are as follows: (a) shows the difference between HQ-SAM and the SumThreshold method; (b) shows the difference between SAM and the SumThreshold method; and (c) shows the difference between HQ-SAM and SAM.}
    \label{comparison-2}
\end{figure*}

\begin{figure*}[hbt!]
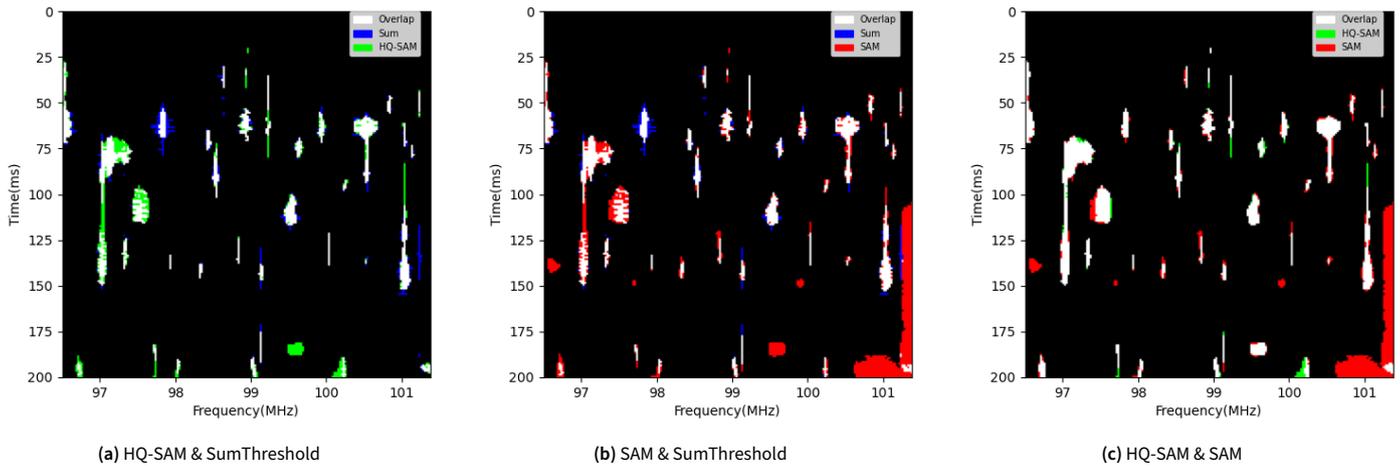

    \centering
    
    \begin{subfigure}{0.3\textwidth}
        \centering
        \includegraphics[width=1.15\linewidth]{fig/RFIresultcomparison/hqsam-sum-2_14600.pdf}
        \caption{HQ-SAM \& SumThreshold}
      \label{comparison-3-b}
    \end{subfigure}
 \hfill
    \begin{subfigure}{0.3\textwidth}
        \centering
        \includegraphics[width=1.15\linewidth]{fig/RFIresultcomparison/sam-sum-2_14600.pdf}
        \caption{SAM \& SumThreshold}
   \label{comparison-3-c}
    \end{subfigure}
\hfill
    \begin{subfigure}{0.3\textwidth}
        \centering
        \includegraphics[width=1.15\linewidth]{fig/RFIresultcomparison/sam-samhq-2_14600.pdf}
        \caption{HQ-SAM \& SAM}
     \label{comparison-3-d}
    \end{subfigure}
    \caption{ The same as in Fig.~\ref{comparison-2}, but comparing the results from the three methods for Fig.~\ref{Fig3:transient narrowband RFI}}
    \label{comparison-3}
\end{figure*}

We plot the detection results of the three methods for Fig.~\ref{Fig2:continuous narrowband RFI} and Fig.~\ref{Fig3:transient narrowband RFI} pairwise on the same layer, using different colors to represent the differences in the masks obtained by each method. As shown in Fig.~\ref{comparison-2-b} and Fig.~\ref{comparison-3-b}, HQ-SAM demonstrates similar performance to the SumThreshold method when utilized on real observational data from the 21CMA, especially in finding as many RFIs as possible. Fig.~\ref{comparison-2-c}, Fig.~\ref{comparison-2-d}, Fig.~\ref{comparison-3-c}, and Fig.~\ref{comparison-3-d} show that SAM outputs more noise than HQ-SAM and the SumThreshold method.

\section{Supplement to RFI Recognition Speed }
\label{appendix speed}

For SAM and HQ-SAM, their runtime are comparable to that of the SumThreshold method when $\rm{points\_per\_side = 40}$ and $\rm{points\_per\_side = 34}$. Under these parameter conditions, the evaluation metrics for Type A and Type B mock RFI recognition tasks can be found in Table~\ref{appendix table1} and Table~\ref{appendix table2}. Table~\ref{appendix table2} also includes the evaluation metrics for SAM and HQ-SAM when they achieve F1 Score roughly comparable to the SumThreshold method at $\rm{points\_per\_side = 64}$ and $\rm{points\_per\_side = 40}$, respectively.

\begin{table}[hbt!]

\caption{Accuracy, Recall, Precision, and F1 Score of SAM and HQ-SAM for the Type A mock RFI recognition task when the three methods have comparable runtime.\\  \textit{Note: pps refers to points\_per\_side.}}
\label{appendix table1}
\begin{tabular}{cccccc}
\toprule
\headrow  & pps & Accuracy   & Recall & Precision & F1 Score \\
\midrule
HQ-SAM&34 & 99.15\% &  52.27\%   & 83.89\%  &64.41\%   \\
\midrule
SAM &40 & 99.21\% & 64.60\%     & 80.78\% &  71.79\%          \\
\bottomrule
\end{tabular}
\end{table}

\begin{table}[hbt!]

\caption{Accuracy, Recall, Precision, and F1 Score of SAM and HQ-SAM for the Type B mock RFI recognition task when the three methods have comparable runtime or F1 Score. \\  \textit{Note: pps refers to points\_per\_side.}}
\label{appendix table2}
\begin{tabular}{cccccc}
\toprule
\headrow  & pps & Accuracy   & Recall & Precision & F1 Score \\
\midrule
HQ-SAM&34 & 99.09\% &  68.08\%   & 90.41\%  &77.67\%   \\
\midrule
HQ-SAM &40 & 99.28\% &78.98\% &91.56\% &84.81\%\\
\midrule
SAM &40 & 98.96\% & 76.83\%     & 82.11\% &  79.38\%          \\
\midrule
SAM &64  &99.20\%   &   89.72\%     & 82.59\%     & 86.01\%           \\
\bottomrule
\end{tabular}
\end{table}

\section{Supplement to SAM's Dependency of Color  }

When the simulated narrowband RFI in Section~\ref{sect:RFI simulation} is converted from RGB format to pseudocolor grayscale images, the detection results of SAM and HQ-SAM are shown in Table~\ref{color table1} and Table~\ref{color table2}. By comparing the evaluation metrics in Table~\ref{Tab1} and Table~\ref{Tab2}, we observe that Recall remain largely unchanged, while Precision show a significant decline (generally over 30\%). This is because when recognizing pseudocolor grayscale images, the models tend to produce coarser boundaries for the detected narrowband RFI compared to images in colormap 'viridis' format, flagging more non-RFI signals in the surrounding areas.

\begin{table}[hbt!]
\caption{Accuracy, Recall, Precision, and F1 Score of SAM and HQ-SAM for the Type A mock RFI recognition task when the images are in pseudocolor grayscale format.}
\label{color table1}
\begin{tabular}{ccccc}
\toprule
\headrow  & Accuracy   & Recall & Precision & F1 Score \\
\midrule
HQ-SAM  & 98.62\% & 92.65\%     & 54.10\% &  68.31\%          \\
\midrule
SAM & 97.75\% &  97.12\%   & 38.67\%  &55.32\%   \\
\bottomrule
\end{tabular}
\end{table}

\begin{table}[hbt!]
\caption{Accuracy, Recall, Precision, and F1 Score of SAM and HQ-SAM for the Type B mock RFI recognition task when the images are in pseudocolor grayscale format.}
\label{color table2}
\begin{tabular}{ccccc}
\toprule
\headrow  & Accuracy   & Recall & Precision & F1 Score \\
\midrule
HQ-SAM  & 98.52\% & 92.65\%     & 66.61\% &  77.96\%          \\
\midrule
SAM & 97.33\% &  96.73\%   & 49.35\%  &65.36\%  \\
\bottomrule
\end{tabular}
\end{table}

\section{Supplement to Sular Radio Burst Detection Results }
\label{SRB detection results appendix}

\begin{figure*}[hbt!]
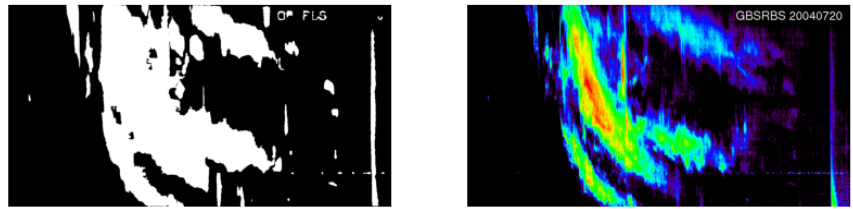
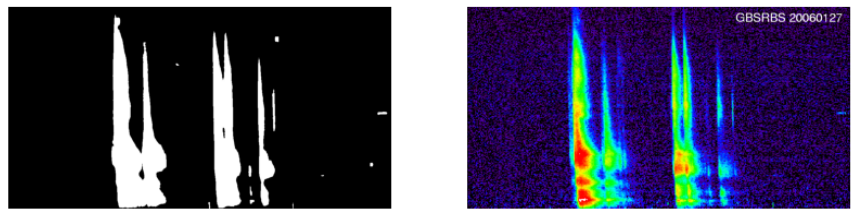
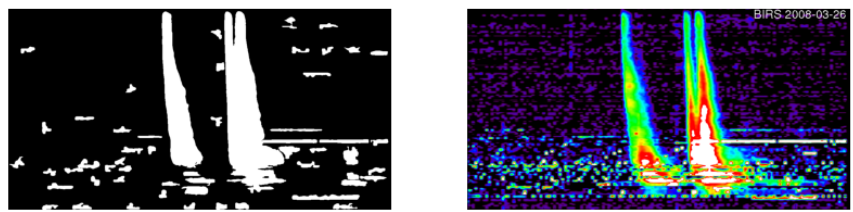
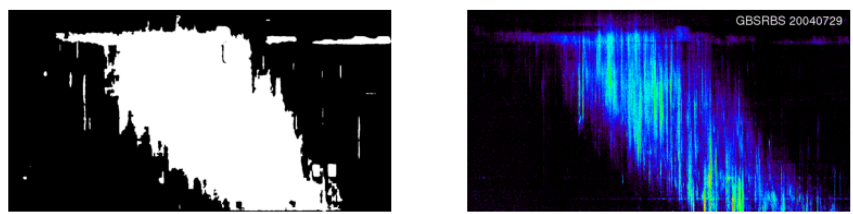
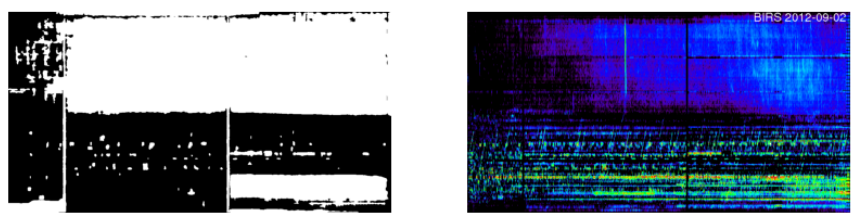

    \centering
    \begin{subfigure}{0.7\textwidth}
        \centering
        \includegraphics[width=0.9\linewidth]{fig/SRB/SRBresults/success/typeII_1.pdf}
        \caption{successful detection of Type II SRB}
      
    \end{subfigure}
 \hspace{1ex}
    \begin{subfigure}{0.7\textwidth}
        \centering
        \includegraphics[width=0.9\linewidth]{fig/SRB/SRBresults/success/typeII_2.pdf}
        \caption{successful detection of Type II SRB}      
    \end{subfigure}
 
    \begin{subfigure}{0.7\textwidth}
        \centering
        \includegraphics[width=0.9\linewidth]{fig/SRB/SRBresults/success/typeIII_1.pdf}
        \caption{successful detection of Type III SRB}     
    \end{subfigure}
  
    \begin{subfigure}{0.7\textwidth}
        \centering
        \includegraphics[width=0.9\linewidth]{fig/SRB/SRBresults/success/typeIII_2.pdf}
        \caption{successful detection of Type III SRB}
      
    \end{subfigure}
    \begin{subfigure}{0.7\textwidth}
        \centering
        \includegraphics[width=0.9\linewidth]{fig/SRB/SRBresults/success/typeIV_1.pdf}
        \caption{successful detection of Type IV SRB}
       
    \end{subfigure}
   
    \begin{subfigure}{0.7\textwidth}
        \centering
        \includegraphics[width=0.9\linewidth]{fig/SRB/SRBresults/success/typeIV_2.pdf}
        \caption{successful detection of Type IV SRB}
      
    \end{subfigure}
    
    \caption{More examples of successful detection for separat Type II-IV SRBs are shown here.  }
    \label{success supplement1}
\end{figure*}

\begin{figure*}[hbt!]
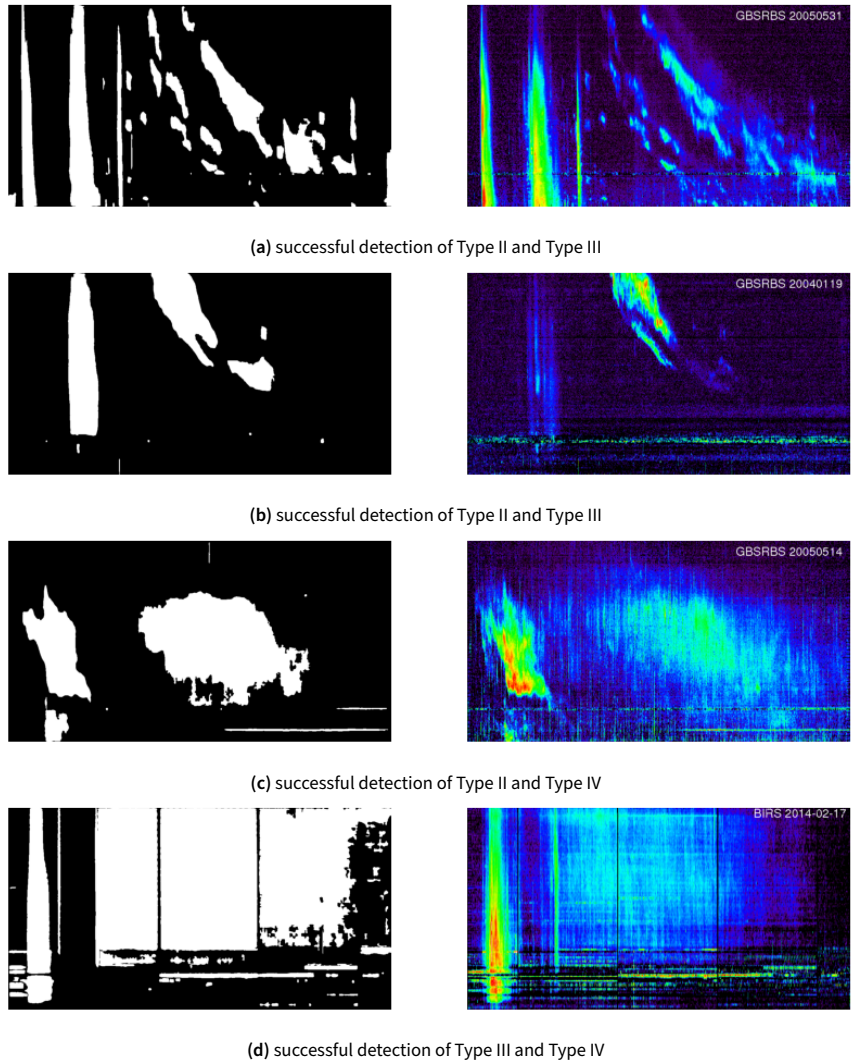

    \centering

    \begin{subfigure}{0.7\textwidth}
        \centering
        \includegraphics[width=0.9\linewidth]{fig/SRB/SRBresults/success/II_III_1.pdf}
        \caption{successful detection of Type II and Type III}

    \end{subfigure}

        \begin{subfigure}{0.7\textwidth}
        \centering
        \includegraphics[width=0.9\linewidth]{fig/SRB/SRBresults/success/II_III_2.pdf}
        \caption{successful detection of Type II and Type III}
   
    \end{subfigure}

        \begin{subfigure}{0.7\textwidth}
        \centering
        \includegraphics[width=0.9\linewidth]{fig/SRB/SRBresults/success/II_IV.pdf}
        \caption{successful detection of Type II and Type IV}
     
    \end{subfigure}

        \begin{subfigure}{0.7\textwidth}
        \centering
        \includegraphics[width=0.9\linewidth]{fig/SRB/SRBresults/success/III_IV.pdf}
        \caption{successful detection of Type III and Type IV}
     
    \end{subfigure}   
    \caption{Examples of successful detection for pairs of Type II-IV SRBs.}
    \label{success supplement2}
\end{figure*}

\begin{figure*}[hbt!]
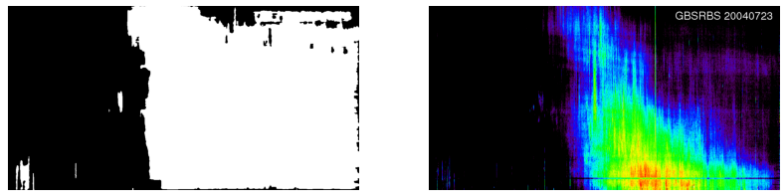
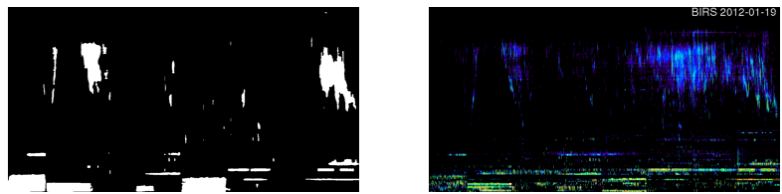
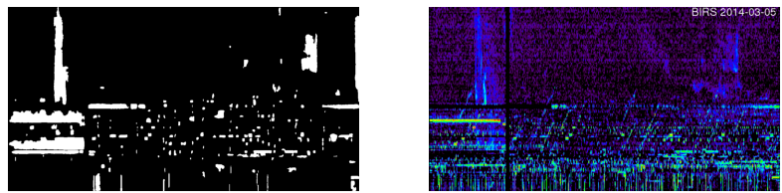
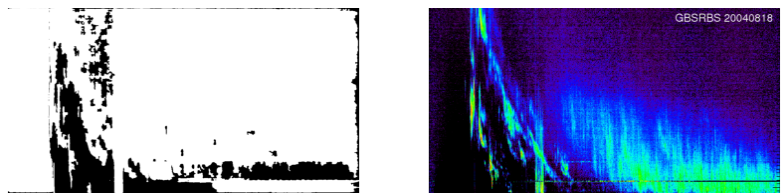
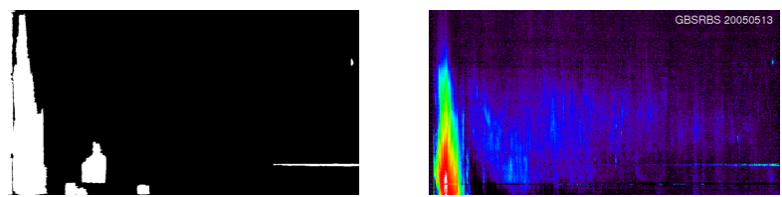

    \centering
    \begin{subfigure}{0.7\textwidth}
        \centering
        \includegraphics[width=0.825\linewidth]{fig/SRB/SRBresults/failure/failII_20040409_200500_BIRS.pdf}
        \caption{failed detection of Type II}
       
    \end{subfigure}
 \hspace{1ex}
    \begin{subfigure}{0.7\textwidth}
        \centering
        \includegraphics[width=0.825\linewidth]{fig/SRB/SRBresults/failure/failIII_20040119_200500_BIRS.pdf}
        \caption{failed detection of Type III}
      
    \end{subfigure}
 
    \begin{subfigure}{0.7\textwidth}
        \centering
        \includegraphics[width=0.825\linewidth]{fig/SRB/SRBresults/failure/failIV_20040723_181500.pdf}
        \caption{failed detection of Type IV}
        \label{large area mask1}
    \end{subfigure}
  
    \begin{subfigure}{0.7\textwidth}
        \centering
        \includegraphics[width=0.825\linewidth]{fig/SRB/SRBresults/failure/failIV_20120119_200500_BIRS.pdf}
        \caption{failed detection of Type IV}
        
    \end{subfigure}

    \begin{subfigure}{0.7\textwidth}
        \centering
        \includegraphics[width=0.825\linewidth]{fig/SRB/SRBresults/failure/failII_III20140305_200500_BIRS.pdf}
        \caption{failed detection of Type II and Type III}
        
    \end{subfigure}
   
    \begin{subfigure}{0.7\textwidth}
        \centering
        \includegraphics[width=0.825\linewidth]{fig/SRB/SRBresults/failure/failIV_II_20040818_173000.pdf}
        \caption{failed detection of Type II and Type IV}
        \label{large area mask2}
    \end{subfigure}
   \begin{subfigure}{0.7\textwidth}
        \centering
        \includegraphics[width=0.825\linewidth]{fig/SRB/SRBresults/failure/failIV_III_20050513_163930.pdf}
        \caption{failed detection of Type III and Type IV}
       
    \end{subfigure}    
    \caption{More examples of failed detection for different types of SRBs are shown here.  }
    \label{failure supplement}
\end{figure*}

More examples of SRB detection results in Section~\ref{HQ-SAM SRB} are presented here (as shown in Fig.~\ref{success supplement1}, Fig.~\ref{success supplement2}, and Fig.~\ref{failure supplement}).

\end{document}